\documentstyle[preprint,aps]{revtex}
\tightenlines
\def\be{\begin{equation}}
\def\ee{\end{equation}}
\def\bea{\begin{eqnarray}}
\def\eea{\end{eqnarray}}
\def\ba{\begin{array}}
\def\ea{\end{array}}
\def\a{\alpha}
\def\b{\beta}

\def\d{\delta}

\def\0{$\Gamma_0$}

\def\o{\omega}
\def\p{\phi}

\def\t{\theta}
\def\tt{\tau}

\def\L{${\cal L}$ }

\begin{document}
\draft

\title{Partition function zeroes of a self-dual Ising model}
  
\author{Wentao T. Lu and F. Y. Wu \\
Department of Physics\\
 Northeastern University, Boston,
Massachusetts 02115}

\maketitle

\begin{abstract}
We consider the Ising model on an $M\times N$ rectangular lattice with 
an asymmetric self-dual boundary condition, and derive a closed-form
expression for its partition function. 
We show that zeroes of the partition function are given by the roots
of a  polynomial equation of  degree $2M-1$,
which trace out
  certain loci in the complex temperature plane.  
Particularly,
it is shown that (a)  real solutions of the   polynomial equations 
always lead to zeroes  on 
 the unit circle and a segment of the negative real axis,
and (b) all temperature zeroes lie on two circles in the limit of $M\to\infty$
for any $N$.
 Closed-form expressions of the  loci as well as the density
of zero distributions in the limit of $N\to\infty$
are derived for  $M=1$ and 2.
In addition, we explain the reason of, and establish the criterion for, 
partition function zeroes of
any self-dual spin model to reside precisely on the unit
circle.  This elucidates a recent  finding  
in the case of the self-dual Potts model.

  \end{abstract}

\vskip 1cm
\pacs{05.50.+q}

\section{Introduction}
One of the most remarkable results in  statistical mechanics is the 
circle theorem of Yang and Lee \cite{yanglee} who in 1952 established   that
zeroes of the  partition function  of the ferromagnetic Ising model   always
lie on the unit circle in the complex $e^{-2L}$ plane, where
$L=H/kT$ is the reduced  magnetic field variable.
In 1964 Fisher \cite{fisher}  pointed out  that it is also
meaningful to consider   zeroes of the Ising partition function
in the complex temperature plane,
in which the zeroes  reside on a boundary 
 at which the
free energy becomes non-analytic. 
Particularly, he showed that temperature zeroes
of the square lattice Ising model lie on two circles in the
thermodynamic limit.  Since then the
consideration of the partition 
function  zeroes has become a power tool in analyzing
lattice models.
The temperature zeroes of
the Ising model \cite{joseph,shrock}
as well as the Potts model \cite{wu} have     been investigated.
However,  studies to this date have been confined mostly to results in the
thermodynamic limit \cite{fisher,joseph} and/or numerical 
analyses \cite{shrock,wu}.
Very little progress has been made for finite lattices,
  except in one dimension \cite{wu1},
in which the locations of zeroes are 
determined in a closed-form and analyzed algebraically.  

 In this paper we  study  the partition function
zeroes of the Ising model for  a finite rectangular 
lattice with an asymmetric self-dual boundary condition.
  This study is  motivated
by  a recently discovered numerical evidence 
in the case of self-dual Potts models \cite{wu} that
many partition function zeroes
  lie precisely on the unit circle. 
  Here, we explain 
why this is the case, and establish more generally the criterion for zeroes
to occur on the unit circle for self-dual lattice models.
Particularly, for the Ising model considered here,
we show  that  all zeroes lie 
on two  circles in the limit of $M\to\infty$ and any $N$.  
We also obtain closed-form expressions for
 the loci of zeroes  for  $M=1,2$,  as well as
the density of zero distribution in the limit of $N\to\infty$.
 
\section{A self-dual rectangular lattice}
Consider a rectangular Ising lattice \L of $M$ rows and $N$ columns with  periodic
boundary conditions in the horizontal direction.  
Let the nearest-neighbor interaction be $-kTK$.  Further, 
introduce a boundary condition that  all
sites on one of the two horizontal boundaries, say, the lowermost one,
interact with an additional site   with the same interaction $-kTK$.
Thus, there are altogether $MN+1$ sites and $2MN$ edges.
Note that the boundary condition is ``asymmetric" in the horizontal and vertical
directions.

Topologically, \L assumes the form of a "wheel" consisting of
$N$ spokes and $M$ concentrate circles.  An example of \L is shown
in fig. 1(a) where the circumference of circle corresponds to
the "horizontal" direction. We point out that the lattice \L is self-dual, which is
 an important premise of our consideration.

The high-temperature expansion of the partition function assumes the form
\be
Z_{M,N} = 2^{MN+1} (\cosh K)^{2MN} G_{\cal L}(z)  \label{part}
\ee
where $z = \tanh K$ and 
\be
G_{\cal L}(z) = 1 + \sum _{{\rm c.p.}\in \>{\cal L}} z^b. \label{graph}
\ee
Here, the summation in (\ref{graph}) is  over all closed polygonal configurations
 that can be   drawn on ${\cal L}$,
and $b$ is the number of  edges contained in each polygonal configuration. 

To facilitate our considerations, we expand the center point   of the wheel
in Fig. 1(a) into a  circle as shown in Fig. 1(b), and associate edge
weights 1 to the $M$ newly added edges.  This transforms the lattice
\L into ${\cal L}^*$.  A moment's reflection \cite{note} shows that we have
the relation
\be
G_{{\cal L}}(z)  ={1\over 2}\> G_{{\cal L}^*}(z). \label{graph1}
\ee
It follows that the partition function (\ref{part}) is computed if we can
evaluate $G_{{\cal L}}^*(z)$.

 The lattice ${\cal L}^*$ is  an $(M+1)\times N$ rectangular lattice
with a periodic boundary condition in the horizontal direction.
All  edges of ${\cal L}^*$ carry the weight $z$ except 
those on  the innermost circle which carry the weight $1$.                                                 
Following Kasteleyn \cite{kast}  we can express $G_{{\cal L}^*}(z) $
as a Pfaffian which is, in turn,  the
square root of a determinant.  The  procedure, which is standard, 
has been described in details by McCoy and Wu \cite{mccoy,mccoy1},
and leads to
  \be 
G_{{\cal L}^*}(z)= \biggl[ \prod _{n=1}^N \det |B_M(\p)|
\biggr]^{1/2}, \label{graph2}
\ee
where ${B_M(\p)}$ is a $4(M+1)\times 4(M+1)$ matrix given by
\be
B_M(\p)=  \pmatrix{B(z) & B_+(z) &        & &  \cr
                   B_-(z) & B(z) & B_+(z) & &  \cr
                       \vdots   &\vdots      &     \ddots& \vdots &\vdots \cr
                          &             & B_-(z) & B(z) & B_+(z) \cr
                          &             &       & B_-(z) & B(1) \cr}.
\label{matrixb}
\ee 
  Here, the product in (\ref{graph2}) is taken over the $N$ values
\be
\phi = (2n-1)\pi/N,\hskip 1cm n=1,2,\cdots N, \label{phi}
\ee
and $B(z)$ and $B_\pm(z)$ are $4\times 4$ matrices given by
(\ref{bm}) in the Appendix.  For completeness, an 
 outline of the derivation
of (\ref{graph2}) is given in the Appendix.  Note that $\p \not= 0, \pi$.
The substitution of (\ref{matrixb}) into (\ref{graph1}) and (\ref{part}) now
expresses the partition function $Z_{M,N}$ in terms of a $4(M+1)\times 4(M+1)$
determinant.

\section{Evaluation of the partition function}
 In this section we evaluate
the determinant $\det |B_M(\phi)|$ in (\ref{graph2}) using 
an approach somewhat different from that of
 \cite{mccoy,mccoy1},
   and present the solution in a  form more suitable
for our purposes.

For brevity we write $B_M\equiv \det|B_M(\phi)|$.
The determinant $B_M$ is bilinear in the  two non-vanishing elements
$ z$ in $B_+(z)$ in the first row and $-z$ in $B_-(z)$ in the first column
of (\ref{matrixb}).  Explicitly, the  bilinear expansion is 
\be
B_M= \det |B(z)|\cdot B_{M-1} + z [B_M]_{1;6} - z [B_M]_{6;1}
+z^2  [B_M]_{1,6;1,6} \label{b}
\ee
where $  [B_M]_{i;k}$ is the determinant of the matrix
 $B_M(\phi)$ with the $i$th row and $k$th column removed, and 
$  [B_M]_{(ij;k\ell)}$ is similarly defined.  We have the following
lemma whose proof is elementary and will not be given:

{\it Lemma}:  For any  any $m\times n$ matrix  $A_{mn}$ and 
 $(N-m)\times (N-n)$ matrix $C_{N-m,N-n}$,  we have the identity
\be
\det \left|\matrix{A_{mn}&0\cr 0& C_{N-m, N-n} \cr} \right|
= \delta_{m,n} \cdot \det |A_{mn}|\cdot  \det |C_{N-m,N-n}|,\label{lemma}
\ee
where $\d$ is the Kronecker delta function.

It follows from the lemma that we have
\be
 [B_M]_{1;6} =  [B_M]_{6;1}=0, 
\ee
\be
 [B_M]_{1,6;1,6} =\det| B(z)|_{1,1} \cdot  [B_{M-1}]_{2;2} .
 \ee
Writing $D_M \equiv [B_M]_{2,2}$
and evaluating  $\det|B(z)|$ and $[B_M]_{1,6;1,6}$ explicitly,
  (\ref{b}) becomes
\be
B_M =a B_{M-1} +b D_{M-1}, \label{bb}
\ee
with 
\be
a = 1+z^2 - 2z \cos \p, \hskip 1cm b=  2iz^3 \sin \p. \label{ab}
\ee
In a similar fashion we expand $D_M$ and obtain 
\be
D_M = c B_{M-1} +d D_{M-1},\label{dd}
 \ee
with 
\be
c= -2iz\sin\p, \hskip 1cm d= z^3 (1+z^2 +2z\cos \p).  \label{cd}
\ee
Both (\ref{bb}) and (\ref{dd}) hold for any $M\geq 1$ with the boundary
condition 
\be
B_0 = \det |B(1)| = 2-2\cos\p, 
\hskip 1cm D_0 = \det|B_0(1)|_{2,2} = -2i\sin\p. \label{bd0}
\ee
Introducing  generating functions
\be
B(\tt) =\sum_{M=0}^\infty B_M \tt^M, \hskip 1cm
D(\tt) =\sum_{M=0}^\infty D_M \tt^M  \label{gen}
\ee
and  multiplying both sides of (\ref{bb}) and (\ref{dd}) by $\tt^M$,
followed by  summing over
 from $M=1$ to $\infty$, one obtains
\bea
B(\tt) -B_0 &=& \tt[ a B(\tt) + b D(\tt)] \nonumber \\
D(\tt) -D_0 &=& \tt[ c B(\tt) + d D(\tt)].
\eea
This yields
\be
B(\tt) ={{B_0(1-d\tt)+D_0 b\tt}\over {1-(a+d)\tt +(ad-bc)\tt^2}}. \label{bfunction}
\ee
It is then a simple matter to obtain an explicit expression for $B_M$ by
 expanding the right-hand side of (\ref{bfunction})
and comparing with  (\ref{gen}),
after first carrying out a partial fraction.
 After some algebra, this leads to the expression
\be
B_M = (2-2\cos\p)[z(1-z^2)]^M \biggl[
{{\sinh[(M+1)t(\p)] - s^{-1} \sinh [Mt(\p)]}\over
{\sinh t(\p)}}\biggr] ,\label{bdet}
\ee
where
\bea
&&2\cosh t(\p) =s+s^{-1} - y 
 \label{ss} \\
&&s \equiv s(K) = e^{2K}\coth K, \hskip 1cm  y=2\>(1+\cos \p). 
          \label{ss1}
\eea
Substituting (\ref{bdet}) into (\ref{graph2}) and (\ref{graph1}), we then
obtain from (\ref{part}) the following closed-form expression for the 
partition function:
\be
Z_{M,N} =2\>
\bigl[2\sinh 2K\bigr]^{MN/2}\> \prod_{n=1}^N \biggl[
{{\sinh[(M+1)t(\p)] - s^{-1} \sinh [Mt(\p)]}\over
{\sinh t(\p)}}\biggr]^{1/2} . \label{part1}
\ee
 Here, use has been made of the identity \cite{gr}
\be
\prod_{n=1}^N \bigg[2-2\cos \biggl({{2n-1} \over N} \pi \biggr)\biggr] =4.
\ee
We remark that we  find also
\be
D_M= -2i[z(1-z^2)]^{M} {{\sin \p}\over {\sinh t(\p)}}
\biggl[ \sinh [(M+1)t(\p)] +s(-K) \sinh [Mt(\p)]
\biggr] .
\ee

\section{zeroes of the partition function}
>From the closed-form expression (\ref{part1}) 
we find 
 the partition function to vanish  at
\be
s={{\sinh[Mt(\p)]} \over{ \sinh [(M+1)t(\p)]}}, 
  \label{zeros}
\ee
a result 
which holds for all $M\geq 1$ and  $\p$ equal to the $N$ values given by (\ref{phi}).
Note that, since only $\cos \p$ appears
in the expression,  distinct zeroes are obtained by using only 
$n=1,2,\cdots,[(N+1)/2]$, where
 $[(N+1)/2]$ is the integer part of $(N+1)/2$,
equal  to $N/2$ for $N=$ even  and $(N+1)/2$ for $n=$ odd.
 It should  also be noted that the numerator and denominator of (\ref{zeros})
 contain a common
factor $\sinh t(\p)$ which cancels, resulting in  polynomials of 
$2\cosh t(\p)$, or $s+s^{-1}-y$,
of degrees $M-1$ and $M$, respectively. 
Equating both sides of (\ref{zeros}) after multiplying by the denominator,
one finds  that the two  terms of the order of
$s^{-M}$  cancel, so that
   (\ref{zeros})
assumes the form
 \be
P_{2M-1}(s)=0,\hskip 2cm n=1,2,\cdots,\biggl[{{N+1}\over 2}\biggr], \label{p}
\ee
where $P_{2M-1}(s)$ is a polynomial of
degree $2M-1$  in $s$.  Explicitly,  we find 
\bea
&& 2\cosh t(\p) - s^{-1} =0, \hskip 6.25cm M=1 \nonumber \\
&& 4\cosh^2 t(\p)  - 2 s^{-1}\cosh t(\p) -1=0, \hskip 3.5cm M=2 \nonumber \\
&&8\cosh^3 t(\p)-4s^{-1} \cosh^2 t(\p) - 4 \cosh t(\p) +s^{-1} =0, \hskip 0.6cm M=3,
\eea
which yield, after using (\ref{ss}),
\bea
&& s-y =0, \hskip 7.6cm M=1 \label{root1}\\
&&s^3 -2ys^2 +y^2 s -y=0, \hskip 5cm M=2 \label{root2}\\
&&s^5 -3ys^4 + 3y^2 s^3 -(2y+y^3) s^2  +2y^2s -y=0, \hskip 1cm M=3.\label{root3}
\eea
Note that  no
closed-form expression exists for 
the roots   for $M\geq 3$.

Guided by results of a recent study \cite{wu}, we now introduce the variable
 \be
x=(e^{2K}-1)/\sqrt 2 \label{xdef}
\ee
so that from (\ref{ss1}) we have
\be
s=\sqrt  2 (x+x^{-1})+3. \label{s}
\ee
This says that  if $x$ is a root of (\ref{p}), then $x^{-1}$ is also a root.
  Furthermore, since $|x+x^{-1}| \leq 2$ if, and only if,  $x$ is
on the unit circle $|x|=1$,
   all real solutions
of (\ref{p})  in the regime
\be
3-2\sqrt 2 \leq s \leq 3+2\sqrt 2 \label{sbound}
\ee
lead to solutions on $|x|=1$.  
In fact, more generally,
we have the following theorem:

{\it Theorem 1}:
Let $P_L(x)$ be 
a polynomial    of degree $L$ in $x$  satisfying the reciprocal relation
\be
P_L(x) = x^{L}P_L(x^{-1}). \label{recip}
\ee
 Then  we have
(i) $P_L(x)$ can be rewritten as
a polynomial in a new variable $w$ with
\bea
P_L(x) &=& x^{L/2} Q_{L/2} (w) , \hskip 1cm
        w=x+x^{-1}, \hskip 1.3cm L = {\rm even} \nonumber \\
       &=& x^{L/2} Q_{L} (w), \hskip 1.25cm w= x^{1/2}+x^{-1/2},
\hskip .5cm L = {\rm odd},
\eea
where $Q_L(w)$ is a polynomial of degree $L$  in $w$.

(ii) 
Roots of $P_L(x)=0$ are on the unit circle $|x|=1$ if,
and only if, the corresponding 
roots of the polynomial $Q(w)=0$ are real  and
in the regime $-2\leq w \leq 2$.

\noindent
Proof:

Property (i) follows from the elementary facts that (a)
the reciprocal relation (\ref{recip}) implies  $x$ and $x^{-1}$ appear 
symmetrically in the $Q$ polynomials in the combination of
$X_n=x^n+x^{-n}$, where $n$ is  an integer for $L$ even 
and a half-odd integer for $L$ odd, and (b) $X_n$
 is a polynomial of 
$w$. Property (ii) follows immediately. Q.E.D.

A consequence of Theorem 1 is that
zeroes of the partition function of self-dual spin models,
which by definition satisfy the reciprocal relation (\ref{recip})
in an appropriate dual variable $x$,
reside precisely on the unit circle $|x|=1$
 for all real roots of the 
$Q$ polynomial  in $|w|\leq 2$.
This explains the numerical findings of \cite{wu} that 
  a large number of   zeroes of self-dual Potts partition functions
are  located  on the unit circle.

For the self-dual Ising model on ${\cal L}$, we have further:

{\it Theorem 2}: 
(i) All real roots of the $[(N+1)/2]$ polynomial
equations $P_{2M-1}(s)=0$  are bounded by
\be
-1< s_{\rm min} \leq s \leq s_{\rm max}< 3+2\sqrt 2,
\ee
where the bounds $s_{\rm min}$ and $s_{\rm max}$ depend on  $M$.

\noindent
(ii) Zeroes in the $x$-plane corresponding to real $s$
 lie  either on the negative real axis 
\be
-(\sqrt 2+1)<x_-(s_{\rm min}) \leq x \leq x_+(s_{\rm min})
<-(\sqrt 2-1)<0, \hskip .5cm 
 s_{\rm min}\leq s \leq 3-2\sqrt 2,  \label{x}
\ee
or on the horseshoe segment $\t_0$ to $2\pi - \t_0$ on the unit circle
\be
|x|=1, \hskip 3.6cm   3-2\sqrt 2 \leq s \leq s_{\rm max}. \label{xcircle}
\ee
Here, $x_+(s) x_-(s) =1$ with
\bea
x_\pm (s)& =&(s-3\pm \sqrt{s^2 -6s +1})/2\sqrt 2, \nonumber \\
\t_0 &=& \cos^{-1} [(s_{\rm max} -3)/2\sqrt 2]. \label{cc}
\eea
Note that the horseshoe segment becomes a circle ($\t_0=0$) if $s_{\rm max} = 3+2\sqrt 2$.

\noindent
(iii) Zeroes in the $x$-plane corresponding to $|s|=1$ lie on the circle
$|x+\sqrt 2| =1$.

\noindent
Proof:

(i) We first show that no root exists for $s>s_0\equiv 3+2\sqrt 2$.  
Assume the contrary, namely, a solution $s>s_0$ exists.  Then by
(\ref{ss}) we have
\bea
\cosh t(\p) &=& {1\over 2}(s+s^{-1}) -(1+\cos\p) \nonumber \\
  &\geq &  {1\over 2}(s_0+s_0^{-1})-(1+\cos\p) \nonumber \\
 &=& 2+\cos\p \nonumber \\
&>&1.
\eea
 It follow that $t(\p)$ is real
and  (\ref{zeros}) gives $s<1$ 
 which contradicts the assumption of $s>s_0=3+2\sqrt 2$. Hence no 
real root
exists for $s>s_0$.

Likewise, for $s<0$ solution of (\ref{p}) we deduce
\be
\cosh t(\p) <-1
\ee
for which $t(\p)$ must assume the form
\be
t(\p) = r + i(2n+1)\pi,
\ee
where $r$ is the real part and $n$ is an integer.  Then (\ref{zeros}) leads to
\be
s= {{-\sinh Mr}\over { \sinh (M+1)r} },
\ee
showing that  negative solutions of (\ref{p})
always lie in the interval $(-1, 0)$.  This completes the proof of (i).

(ii)
For $-1<s_{\rm min}<s\leq 3-2\sqrt 2$,  one solves 
$x=x_\pm(s)  $ from (\ref{s}).
This leads to (\ref{x}) where we have used
$x_\pm(-1) = -(\sqrt 2\mp 1)$.  For $3-2\sqrt 2 \leq s \leq s_{\rm max}$, one writes
$x=e^{i\t}$ and solves for $\t$ from (\ref{s}).  This leads to the
result (\ref{xcircle}) with $\t_0$ given by (\ref{cc}).

(iii) It is straightforward to show that $|s|=1$ 
by substituting  $x=-\sqrt 2 +e^{i\b}$ into (\ref{s}) for any real $\b$.
 Q.E.D.

\section{Loci of zeroes}
The loci of zeroes can be 
explicitly determined for $M=1,2$ and in the limit of $M\to \infty$.
Consider first the case of $M=1$.
 
For $M=1$ (\ref{root1}) gives 
\be
s = 2(1+\cos \p)  ,\label{m1}
\ee
so that $s_{\rm min}= 0$ and $s_{\rm max}=4$.
 It follows that the zeroes are either
on the line segment $x=[-\sqrt 2, -1/\sqrt 2]$ 
 or on the horseshoe segment of  $|x|=1$ 
with  $\t_0 = \cos^{-1} (1/2 \sqrt 2) = 57.3^o$.
A plot of the loci obtained for $N=4,000$ is shown in Fig. 2.
 In the limit of large $N$,
 the zeroes are distributed uniformly in  $\p$ with a density
$\rho(\p)=1/2\pi$.
Combining (\ref{m1}), (\ref{s}) and $x=e^{i\t}$, we have
\be
\cos \p = \sqrt 2 \cos \t -1/2,
\ee
and find that 
the zeroes distribute on the horseshoe segment with a density 
\be
\rho (\t) = {1\over {2\pi}} {{d\p}\over {d\t}} = {1\over {\sqrt 2 \pi}}
{{\sin \t}\over {\sin \p}},
\ee
 which diverges at $\t_0$ as $(\t - \t_0)^{-1/2}$.

For $M=2$  the cubic equation (\ref{root2})  yields three roots
\be
s_n={2y\over 3}+\o^n \root 3\of{Q+\sqrt{\Delta}}+
\o^{2n} \root 3\of{Q-\sqrt{\Delta}},\hskip 1cm n=0,1,2,
\ee
where 
 \be
\o =e^{i2\pi /3}, \hskip 1cm
 Q= {y\over 2} \biggl( 1-{{2y^2}\over {27}}\biggr), \hskip 1cm
  \Delta =  {y\over 4}^2 \biggl( 1-{{4y^2}\over {27}}\biggr).
\ee
Now $s_0$ is always real and assumes the minimum value $s=0$ at $y=0$.
In addition, $s_{1,2}$ are  real for $\Delta<0$
or $3\sqrt 3/2 < y < 4$. It follows that we have $s_{\rm min}=0$.
A more detailed analysis  shows that we have  $s_{\rm max} = 8(1+\cos \a)/3$,
where $\a ={1\over 3}  \cos^{-1}(-5/32)$.
These lead to zeroes in the $x$-plane on the negative real
segment $[-\sqrt 2, -1/\sqrt 2]$ and a horseshoe segment of the
unit circle from $\t_0$ to $2\pi - \t_0$, with  $\t_0=\cos^{-1} 
[(8\cos \a -1)/6\sqrt 2] = 47.7^o$.
A plot of the zeroes for $N=2,000$ is shown in Fig. 3.  
The density of zeroes in the limit of $N\to\infty$, which can be found 
accordingly, is given as  the superposition of the densities
of the three branches.  However, since only one of the 
overlapping densities extends
to $\t_0$, the density diverges at $\t_0$
as $(\t - \t_0)^{-1/2}$.

For $M\geq 3$ the polynomial equation (\ref{p}) can be solved 
only numerically.
We have carried out its numerical solution 
for $3\geq M\geq 8$ and determined the corresponding values
 of $\t_0$. 
The result of $M=3, N=400$ is shown in Fig. 4.
We also show in Fig. 5 the trend of $\t_0\to 0$
as $M\to \infty$.  

 In the limit of $M\to \infty$, we note that
for $M$ large we must have $|e^{t(\p)}|=1$.  This is so since, otherwise,  
without the loss of generality we can assume $|e^{t(\p)}|>1$
and (\ref{zeros})  yields $s \sim e^{-t(\p)}$  which contradicts  
 (\ref{ss}) (except at $\p=\pi$).  Thus we  write $t(\p) = i \a(\p)$
where $\a(\p)$ is real.  Then, (\ref{ss}) becomes
\be
 s + s^{-1} = 2 \cos \a +2(1+\cos \p)= {\rm real} \label{bound}
\ee
leading to the bound
$-2 \leq s+s^{-1} \leq 6$.  
For $-2\leq s+s^{-1} \leq 2$ we must have $|s|=1$ 
 which implies,
by Theorem 2(iii), 
that all zeroes are on the circle $|x+\sqrt 2 | =1$.
 For $2\leq s+s^{-1} \leq 6$, the bound is equivalent to the bound (\ref{sbound}) on $s$
and, by Theorem 1(ii), all zeroes are
on the unit circle
 $|x|=1$.  
We therefore conclude that in the limit of $M\to \infty$ 
all zeroes are on the two circles $|x|=1$ and $|x+\sqrt 2 | =1$.
Indeed, this coincides with the finding of
Fisher \cite{fisher} who showed that, in the limit of both $M, N\to \infty$,
the partition function 
zeroes are on the two equivalent circles $|\tanh K \pm 1|= \sqrt 2$
\cite{fisher1}.
However, our result is more  general and holds for 
$M\to \infty$ and any $N$.

Finally,  we remark that our results  confirm the conjecture 
put forth in \cite{wu},
that  all zeroes in the ${\rm Re}\>\> (x) >0$ half plane are on the
unit circle $|x|=1$.  But a generally
proof of this conjecture is still lacking at this point.

\section{Discussions}
We have obtained the closed-form expression of the partition function of
an $M\times N$ rectangular Ising lattice with an asymmetric boundary condition,
and evaluated its zeroes.  It was found that the 
zeroes are given by the roots of the polynomial equation (\ref{p}).
 Particularly, we established that
 real roots of (\ref{p})
lead to zeroes either on the unit circle $|x|=1$ or within the segment
$[-(\sqrt 2 +1), -(\sqrt 2 -1)]$  of the
negative real axis  in the complex
$x$ plane.
We also established that all zeroes lie on two unit circles centered
respectively at $x=0, -\sqrt 2$  in the $M\to\infty$ limit for any $N$,
where  $x=(e^{2K}-1)/\sqrt 2$.
Numerical results for finite $M$ are also presented.

\section{Acknowledgement}
Work has been supported in part by National Science Foundation Grant
DMR-9614170.
 

\setcounter{section}{0}
\renewcommand{\thesection}{APPENDIX}

\begin{center}
\section{Evaluation of a Pfaffian}
\end{center}

\setcounter{equation}{0}
\renewcommand{\theequation}{\Alph{section}\arabic{equation}}

 For completeness we include in this appendix an outline of 
the steps leading to expressing the graph generating function $G_{{\cal L}^*}(z)$
as  a Pfaffian.
 First, following  standard procedures  \cite{kast} we see that the graph generating function
$G_{{\cal L}^*}(z)$ for ${\cal L}^*$
is equal to the dimer generating function for
 a  "bathroom-tile" lattice constructed from ${\cal L}^*$
as shown in Fig. 6.  This leads to the relation 
\be
G_{{\cal L}^*} = {\rm Pf}\> A = \sqrt{ \det |A|},  \label{a1}
\ee
where ${\rm Pf} A$ is 
the Pfaffian of a $4(M+1)N \times 4(M+1)N$ antisymmetric matrix 
\be
A=A_M \otimes I_{N} + C_{M} \otimes H_N - C_{M}^T \otimes H_N^T. \label{a2}
\ee
Here, $I_N$ is the $N\times N$ identity matrix,
 $H_N$ is an $N\times N$  matrix
given by
\be
H_N =  \pmatrix{0 & 1 &        0 & \ldots & 0\cr
                   0 & 0 &1 & \ldots  & 0\cr
                       \vdots   &\vdots      &      \vdots&\ddots &\vdots \cr
                        0  &    0    &    0    & \ldots  & 1 \cr
                          -1&    0     &   0     &  \ldots     & 0 \cr}, 
                          \ee
$A_M$ and $C_M$ are  $4(M+1) \times 4(M+1)$ matrices given by
\bea
 C_{M}&=& \pmatrix{ C(z) &   0     & & & \cr
                    0 & C(z) & & & \cr
                       \vdots      &     \vdots& \ddots&\vdots &\vdots \cr
                                &        & & C(z) & 0 \cr
                                   &        &      & 0 & C(1) \cr},\nonumber \\
 A_{M}&=& \pmatrix{ B(0) &   B_+(z)    & & & \cr
                    B_-(z) & B(0) & B_+(z)& & \cr
                       \vdots      &     \vdots& \ddots&\vdots &\vdots \cr
                                &        &B_-(z) & B(0) & B_+(z) \cr
                                   &        &      & B_-(z)& B(0) \cr}
\eea
with
\bea
 C(z) &=& \pmatrix{ 0&0&0&0\cr 0&0&0&0\cr 0&0&0&z\cr 0&0&0&0\cr},
     \hskip 1cm 
B(z) = \pmatrix{0&1&-1&-1 \cr -1&0&1&-1 \cr 1&-1&0& 1+ze^{i\phi} \cr
                1&1& -(1+ze^{-i\phi})&0 }, \nonumber \\
B_+(z)&=&\pmatrix{ 0& z& 0& 0 \cr 0& 0& 0& 0 \cr 0& 0& 0& 0 \cr 0& 0& 0& 0 },
\hskip 1cm
B_-(z) =\pmatrix{ 0&0& 0& 0 \cr -z& 0& 0& 0 \cr 0& 0& 0& 0 \cr 0& 0& 0& 0 }.\label{bm}
\eea
 Here $H_N^T$ and $C_M^T$ denote, respectively,
 the transpose of $H_N$ and $C_M$.

Now, $H_N$ and $H_N^T$ can be diagonalized simultaneously with 
respective eigenvalues $e^{i\p}$ and $e^{-i\p}$, where
$\p = \pi (2n-1)/N$, $n=1,2,\cdots,N$. We can therefore
replace $H_N$ (resp. $H_N^T$) by a diagonal matrix with diagonal elements 
 $e^{i\p}$ ( resp. $e^{-i\p}$) without affecting the determinant.
  The substitution of (\ref{a2}) into
(\ref{a1}) now leads to (\ref{graph2}).
 
\newpage 
\centerline{\bf Figure Captions}
\bigskip
\noindent
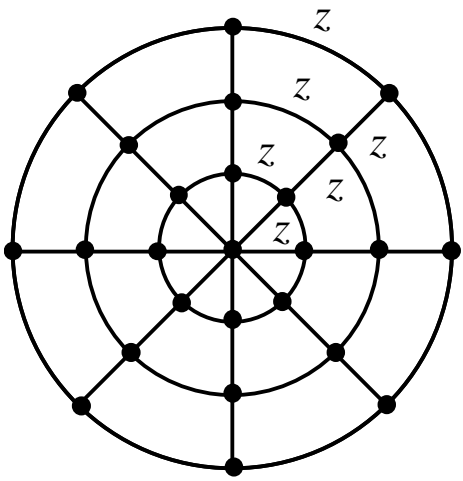
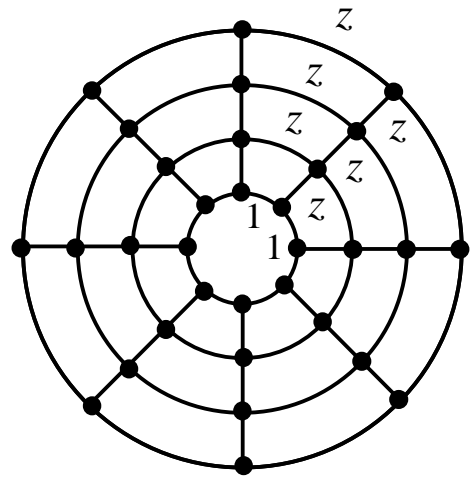
Fig. 1. (a) A $3\times 8$ self-dual lattice ${\cal L}$.  
(b) The $4\times 8$
lattice ${\cal L}^*$ derived from ${\cal L}$. 

\noindent
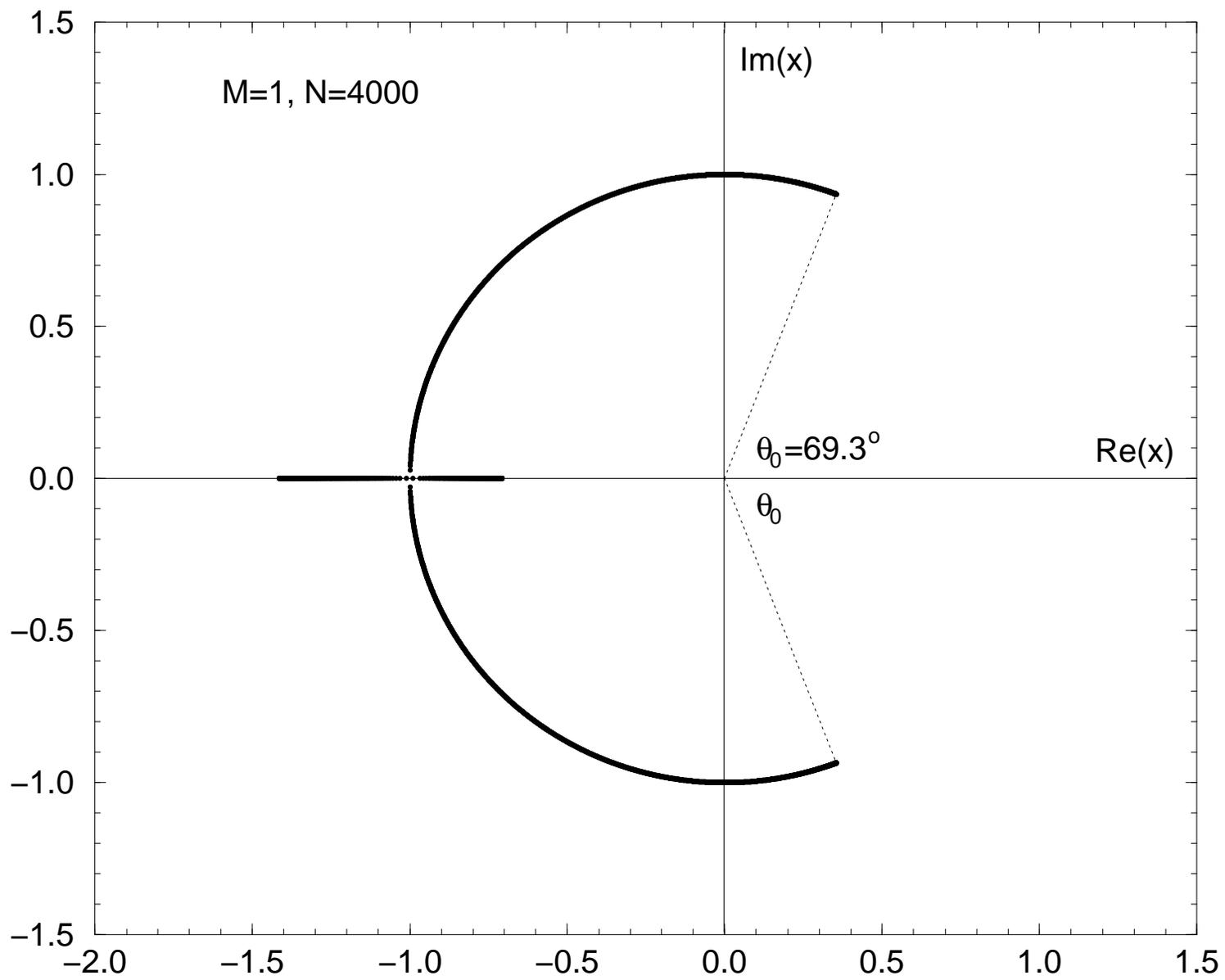
Fig. 2. Loci of zeroes in the $x$ plane for $M=1$, $N=4,000$.

\noindent
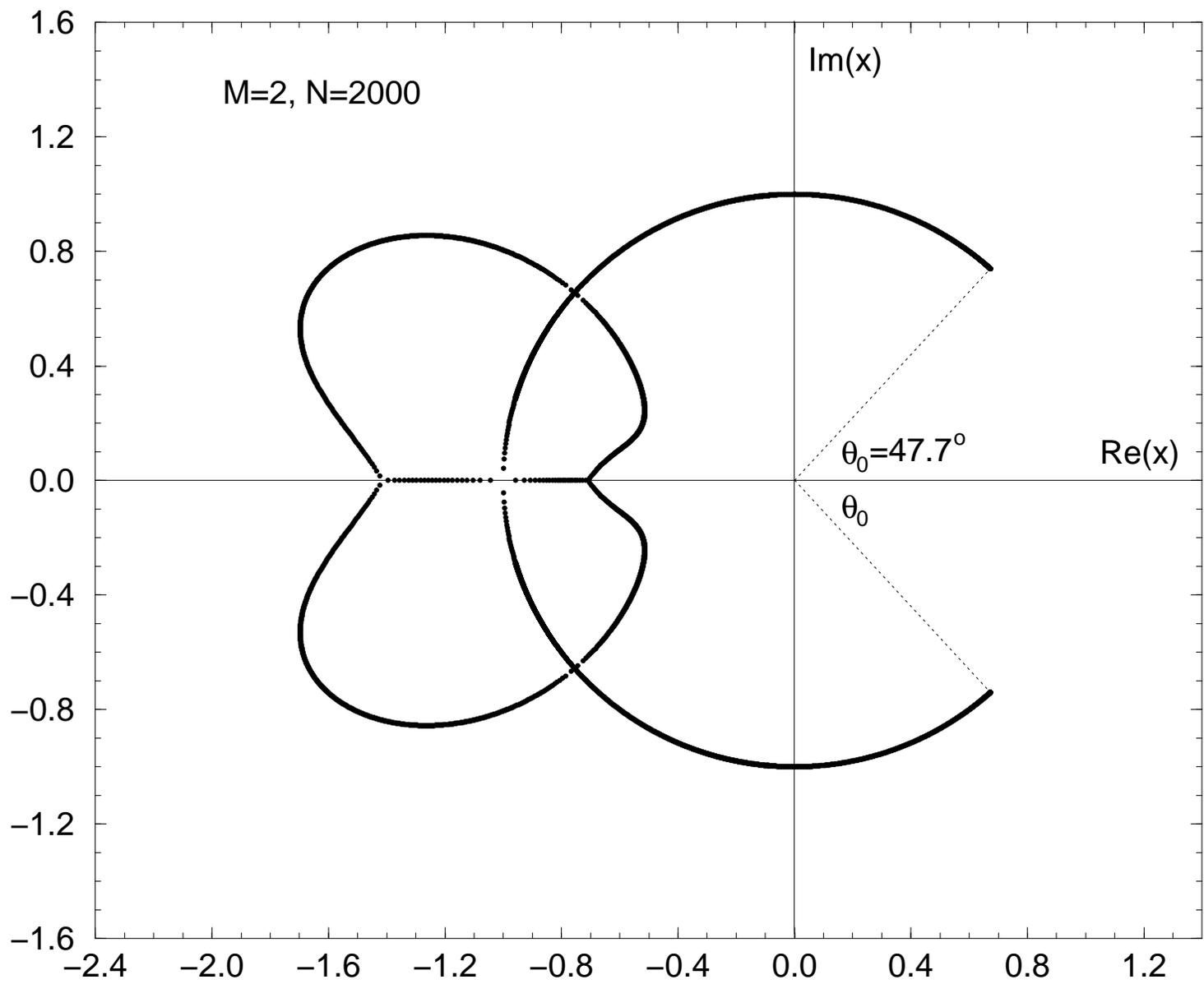
Fig. 3. Loci of zeroes in the $x$ plane for $M=2$, $N=2,000$.

\noindent
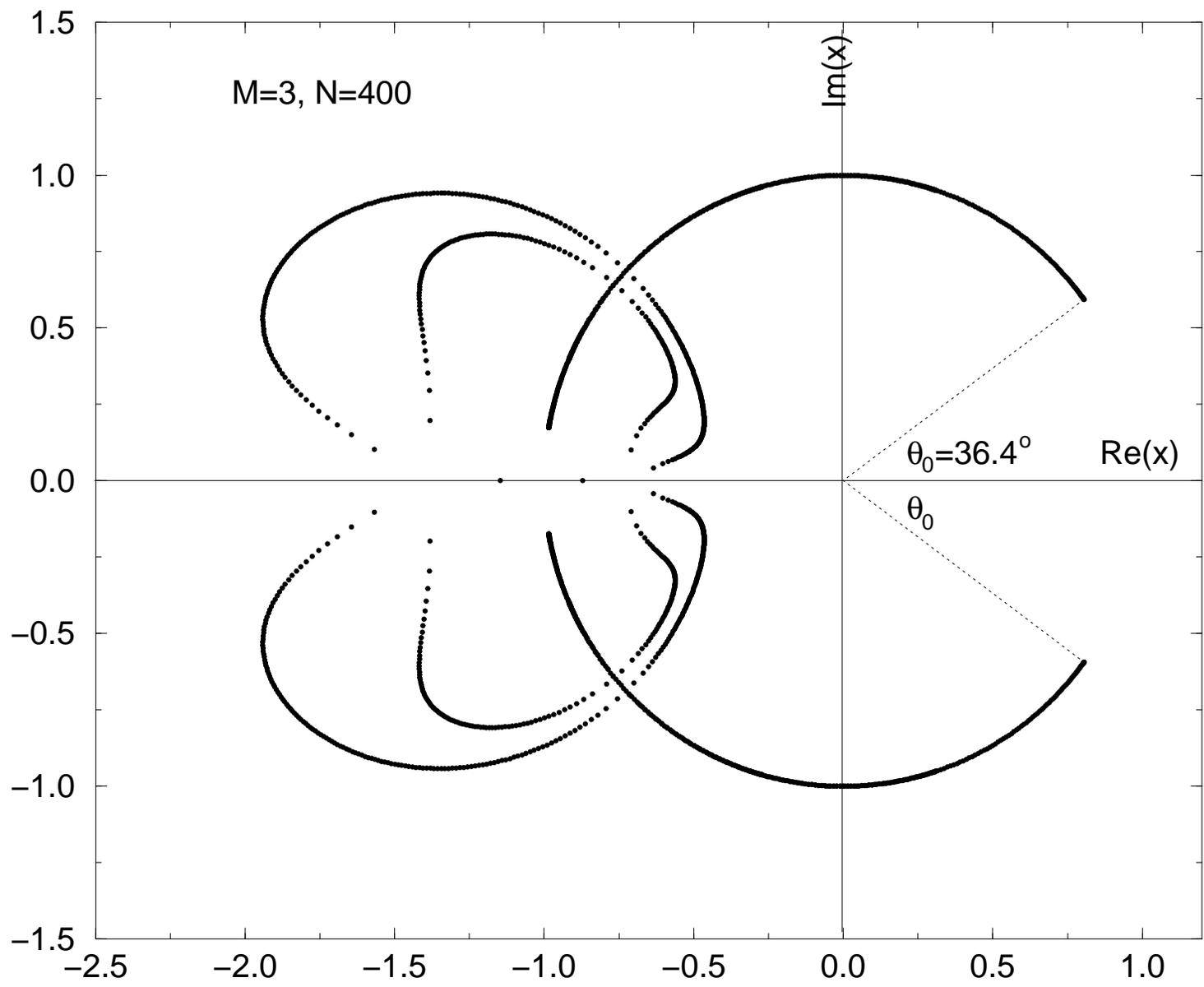
Fig. 4. Loci of zeroes in the $x$ plane for $M=3$, $N=400$.

\noindent
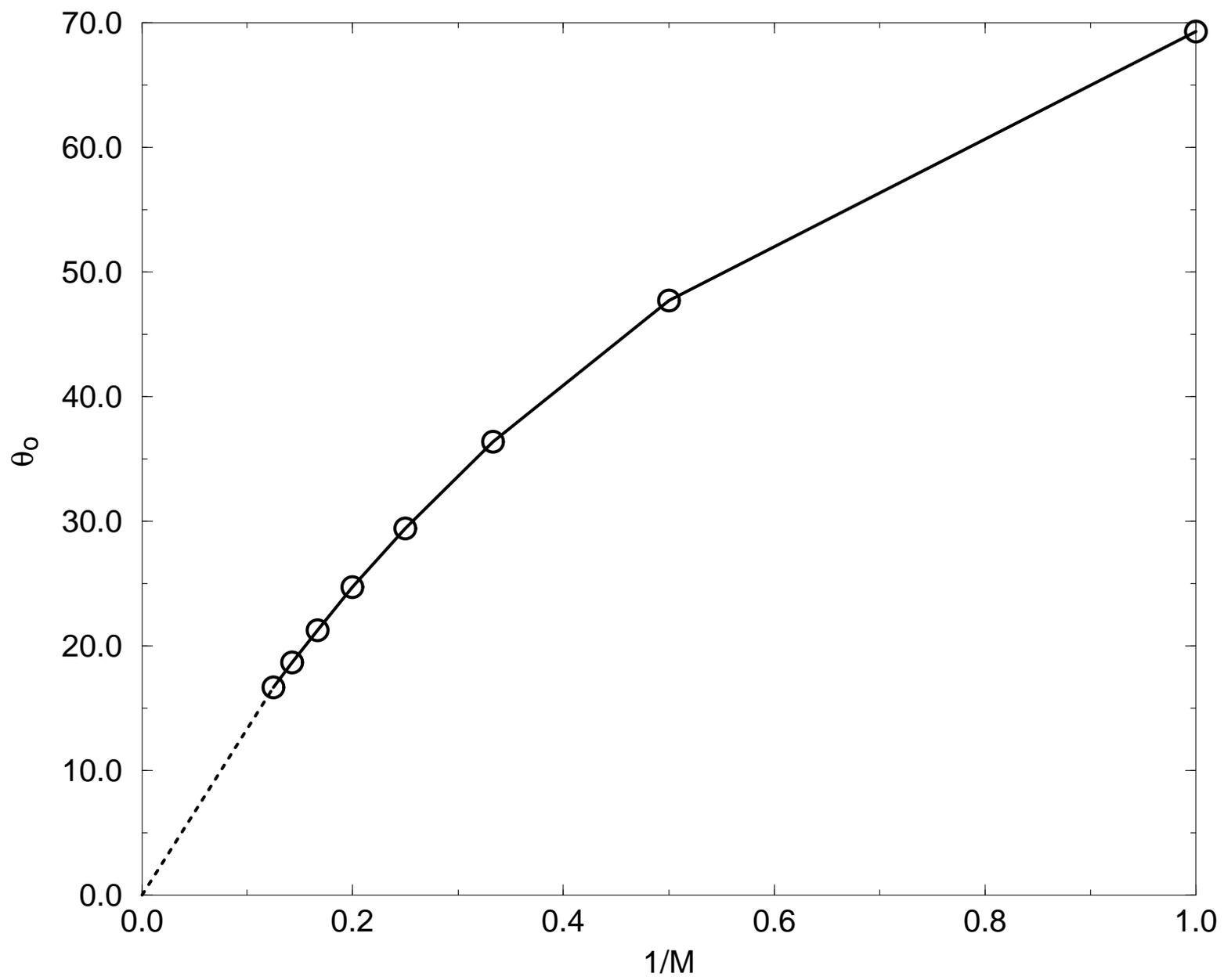
Fig. 5. The dependence of $\t_0$ on $M$.

\noindent
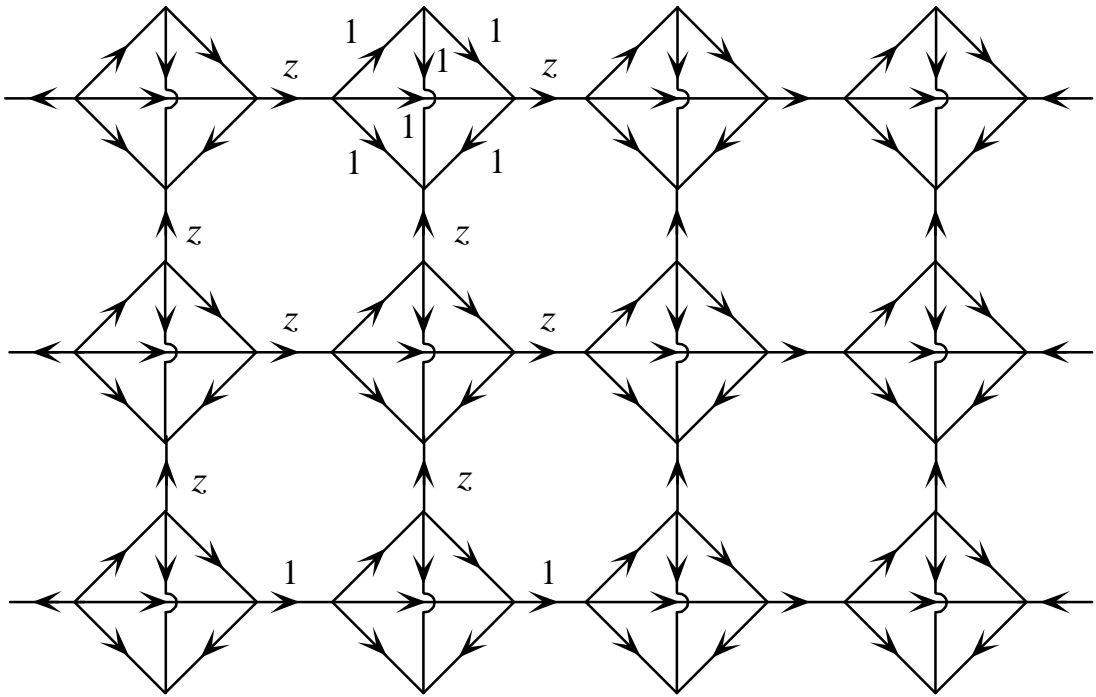
Fig. 6.  The bathroom-tile lattice for a $3\times 4$ 
lattice ${\cal L}^*$.


\begin{thebibliography} {9}
\bibitem{note} Eq. (\ref{graph1}) follows from the fact that to  each
closed polygonal configuration in \L  there exist two closed polygonal
configurations in ${\cal L}^*$, obtained by completing the polygons 
along the innermost circle  in two different
ways.  These two configurations carry the same weight since the edge weight
for the innermost circle is $1$.
\bibitem{fisher1} Note that in comparing the two  results, the caption
of Fig. 19.1 of \cite{fisher} contains a misprint where $v=\tanh(K/2)$
should read $v=\tanh K$.
\end{thebibliography}
\end{document}